\documentclass[aps,twocolumn,superscriptaddress,floatfix,nofootinbib]{revtex4-1}
\usepackage{graphicx,amsmath,amssymb,verbatim,color}
\usepackage{booktabs}
\usepackage{comment}
\usepackage{soul}
\usepackage[dvipsnames]{xcolor}
\usepackage{bm}
\usepackage[utf8]{inputenc}
\usepackage[colorlinks=true,citecolor=blue,linkcolor=blue,urlcolor=blue, backref=false,pdfborder={0 0 0}]{hyperref}
\usepackage{float}
\usepackage{multirow}
\newcommand{\be}{\begin{equation}\begin{gathered}}
\newcommand{\ee}{\end{gathered}\end{equation}} 
\newcommand{\barr}{\begin{eqnarray}}
\newcommand{\earr}{\end{eqnarray}} 
\usepackage{romannum}
\usepackage[normalem]{ulem}

\newcommand{\pk}{\mathcal{P}(k)}
\newcommand{\dd}{\mathrm{d}}

\usepackage{makecell}

\begin{document}

\pagenumbering{arabic}

\title{What it takes to solve the Hubble tension\\through scale-dependent modifications of the primordial power spectrum}
\author{Nanoom Lee}
\email{nanoom.lee@jhu.edu}
\affiliation{Center for Cosmology and Particle Physics, Department of Physics, New York University, New York, New York 10003, USA}
\affiliation{William H. Miller III Department of Physics \& Astronomy, Johns Hopkins University, Baltimore, Maryland 21218, USA}
\author{Matteo Braglia}
\email{mb9289@nyu.edu}
\affiliation{Center for Cosmology and Particle Physics, Department of Physics, New York University, New York, New York 10003, USA} 
\author{Yacine Ali-Ha\"imoud}
\email{yah2@nyu.edu}
\affiliation{Center for Cosmology and Particle Physics, Department of Physics, New York University, New York, New York 10003, USA}

\date{\today} 
\begin{abstract}
	We investigate scale-dependent modifications to the primordial scalar power spectrum as potential solutions to the Hubble tension. We use the Fisher-bias formalism, recently adapted to examine perturbed recombination solutions to the Hubble tension, and extend its range of validity with an iterative method. We first analyze the Planck cosmic microwave background (CMB) anisotropy data, demonstrating the existence of modifications to the primordial power spectrum capable of fully resolving the tension between Planck and SH0ES. As a proof of concept, we interpret these solutions in terms of small, time-dependent variations in the first slow roll parameter or in the sound speed of curvature perturbations during a stage of primordial inflation. However, these solutions are associated with a low total matter density $\Omega_m$, which makes them inconsistent with baryon acoustic oscillations (BAO) and uncalibrated supernovae (SNIa) data. When incorporating additional BOSS and PantheonPlus data, the solutions that reduce the Hubble tension tend to overfit Planck CMB data to compensate for the worsened fit to BAO and SNIa data, making them less compelling. These findings suggest that modifying the primordial power spectrum alone is unlikely to provide a robust resolution to the tension and highlight how the viability of such data-driven solutions depends on the specific datasets considered, emphasizing the role of future high-precision observations in further constraining possible resolutions to the tension.
\end{abstract}
\maketitle

\section{Introduction}

A major challenge in modern cosmology is the persistent discrepancy between different measurements of the Hubble constant, $H_0$, which characterizes the expansion rate of the Universe today. Local measurements, such as those from the SH0ES (Supernovae and $H_0$ for the Equation of State of dark energy) collaboration, determine $H_0$ using the cosmic distance ladder, yielding $H_0 = 73.04 \pm 1.04\,{\rm km}\,{\rm s}^{-1}\,{\rm Mpc}^{-1}$ \cite{Riess:2021jrx}. This measurement is largely independent of the assumed cosmological model. In contrast, the Planck 2018 DR3 analysis \cite{Planck:2018vyg}, which infers $H_0$ from cosmic microwave background (CMB) anisotropies under the assumption of a flat $\Lambda$ cold dark matter ($\Lambda$CDM) model, yields a significantly lower value of $H_0 = 67.36 \pm 0.54\,{\rm km}\,{\rm s}^{-1}\,{\rm Mpc}^{-1}$. This tension, commonly referred to as the Hubble tension\footnote{See Ref.~\cite{Poulin:2024ken}, which reframes this discrepancy as a ``cosmic calibration tension", emphasizing that it extends beyond just the measurement of $H_0$ itself.}, now approaching $5\sigma$, is unlikely to be a statistical fluctuation and may point to unresolved systematics or new physics beyond the standard cosmological model \cite{Verde:2019ivm,DiValentino:2020zio,DiValentino:2021izs,Schoneberg:2021qvd,Shah:2021onj,Abdalla:2022yfr}.

Many attempts to resolve this discrepancy have focused on modifying $\Lambda$CDM to increase the inferred $H_0$ from CMB data. Broadly, these solutions fall into two categories: early-time modifications, which reduce the sound horizon $r_s(z_*)$, and late-time modifications, which change the expansion history after recombination to modify the angular diameter distance $d_A(z_*)$ to recombination. Early-time solutions often involve additional energy components, such as early dark energy (EDE) \cite{Karwal:2016vyq,Poulin:2018cxd,Smith:2019ihp}, modifications to gravity \cite{Rossi:2019lgt,Zumalacarregui:2020cjh,Ballardini:2020iws,Braglia:2020iik,Braglia:2020auw,Adi:2020qqf}, extra relativistic species \cite{Blinov:2020hmc,Schoneberg:2022grr}, or modifications to recombination physics \cite{Hart:2019dxi,Jedamzik:2020krr,Lee:2021bmn,Lee:2022gzh,Lynch:2024hzh,Schoneberg:2024ynd,Mirpoorian:2024fka,Jedamzik:2025cax}. Late-time solutions typically invoke modifications to dark energy \cite{DiValentino:2017zyq,Poulin:2018zxs,Camarena:2021jlr}, but these are tightly constrained by baryon acoustic oscillation (BAO) and supernova data \cite{Keeley:2022ojz}. No single proposal has fully resolved the tension, and some have argued that a combination of early- and late-time modifications may be necessary \cite{Vagnozzi:2023nrq,Poulin:2024ken}.

A different approach is to reconsider the initial conditions set by inflation. The standard assumption of a nearly scale-invariant primordial power spectrum may not be fully accurate, and deviations from scale invariance could affect CMB-inferred cosmological parameters, including $H_0$. This possibility has been explored in a series of studies \cite{Hazra:2018opk,Liu:2019dxr,Keeley:2020rmo,Hazra:2022rdl,Antony:2022ert}, with some evidence suggesting that a modification to $\mathcal{P}(k)$ could increase the inferred $H_0$ from CMB data.

Building on this idea, we apply a model-independent, data-driven approach developed in Lee \textit{et al.}~\cite{Lee:2022gzh} to search for modifications to the primordial power spectrum that could resolve the Hubble tension\footnote{See Refs.~\cite{Mirpoorian:2024fka,Zhou:2025kws} for other applications of this generic approach.}. Our method allows for fully flexible adjustments taking into account degeneracies between modifications to $\mathcal{P}(k)$ and cosmological parameters, ensuring a self-consistent cosmological fit\footnote{We emphasize that our method is different from that of Refs.~\cite{Hazra:2022rdl,Antony:2022ert} where a specific cosmology is assumed when finding modifications to $\pk$.}. Furthermore, we introduce an iterative procedure to explore deviations beyond small perturbations to $\Lambda$CDM, advancing the original method of Lee \textit{et al.}~\cite{Lee:2022gzh}. Finally, while our analysis remains agnostic about the specific mechanism that produced the initial perturbations in the early Universe, we take initial steps toward understanding whether the required modification could arise during a primordial stage of inflationary expansion. Using a method similar to that of~\cite{Durakovic:2019kqq,Raffaelli:2025kew}, we interpret the modification to the primordial power spectrum as a time-dependent correction to the slow-roll (SR) parameters or the sound speed of curvature perturbations in the effective field theory (EFT) of inflationary fluctuations in single-field inflation~\cite{Creminelli:2006xe,Cheung:2007st}.

We present, for the first time, a modification to $\mathcal{P}(k)$ which brings the CMB-inferred $H_0$ into perfect agreement with SH0ES. However, the solution with Planck TT/TE/EE spectra requires a larger baryon abundance, inconsistent with the BBN constraint \cite{Schoneberg:2024ifp}. Furthermore, our solutions increasing $H_0$ generally require a lower total matter density $\Omega_m$, which potentially conflicts with the constraints from CMB lensing, BAO and supernovae observations. We investigate this issue by incorporating these datasets into our analysis. When Planck lensing measurements are added to our baseline, a resolution to the $H_0$ tension still exists, although the fit to the Planck lensing data is significantly worsened compared to the $\Lambda$CDM best-fit. This trend becomes more pronounced when BOSS BAO and PantheonPlus uncalibrated supernovae data are included. In that case, our formalism can only produce solutions which raise $H_0$ to about 72 km/s/Mpc, thereby reducing the tension with SH0ES below the 1$\sigma$ level. However, as in the lensing case, this comes at the cost of a substantially degraded fit to BOSS and PantheonPlus data, compensated by an overfitting of Planck data. Moreover, the required modification to $\mathcal{P}(k)$ in this scenario is highly oscillatory and non-perturbative, making the solution significantly less compelling.

This work provides a systematic exploration of the viability of primordial power spectrum modifications as a solution to the Hubble tension. While our results show that such modifications can alleviate the discrepancy, they also highlight the challenges of reconciling a higher $H_0$ with multiple cosmological datasets. Future high-precision CMB and large-scale structure observations will be crucial to test whether deviations in $\mathcal{P}(k)$ are a viable explanation for the tension or whether alternative new physics is required.

This paper is organized as follows. In Sec.~\ref{sec:method} we describe how we apply the method of Lee \textit{et al.}~\cite{Lee:2022gzh} to modifications of the primordial power spectrum. In Sec.~\ref{sec:data} we describe the cosmological datasets considered in this work. In Sec.~\ref{sec:results} we show our main results, which are data-driven solutions to the Hubble tension with three different sets of data: 1) Planck-T\&E only, 2) Planck-T\&E + Planck lensing, and 3) Planck-T\&E + BOSS BAO + PantheonPlus uncalibrated SN\Romannum{1}a, and then discuss about the implications of our results in Sec.~\ref{sec:implications}, in the context of single-field inflation. We conclude in Sec.~\ref{sec:discussion}.

\section{Methodology}
\label{sec:method}

We apply the method developed by Lee \textit{et al.}~\cite{Lee:2022gzh}, which enables searching for data-driven extensions to the standard $\Lambda$CDM model resulting in desired shifts in cosmological parameters while not worsening the fit to given cosmological datasets. We further advance this method by taking an iterative approach. Our goal is to find modifications to the primordial power spectrum that solve the Hubble tension by increasing the inferred value of the Hubble constant from the CMB data.

Assuming the flat $\Lambda$CDM model, cosmological observables are functions of six cosmological parameters, $\vec{\Omega} \equiv \{\omega_c, \omega_b, h, \tau, \ln(10^{10}A_s), n_s\}$, where $\omega_{c,b}\equiv \rho_{c,b} h^2/\rho_{\rm crit}$ are the physical energy density parameters for CDM and baryon, $h$ is the reduced Hubble constant, $\tau$ is the optical depth to reionization, $A_s$ and $n_s$ are the amplitude and spectral index of the primordial scalar power spectrum. Considering perturbations in some smooth function $f$ as an extension to the standard $\Lambda$CDM model, we solve the following constrained optimization problem, as in Lee \textit{et al.}~\cite{Lee:2022gzh}:
\be
\textrm{minimize}(|| \Delta f(\xi) ||^2)  \ \ \textrm{with} \ \begin{cases} H_0^{\rm BF}[\Delta f(\xi)] = H_0^{\rm target}, \\[5pt]
\Delta \chi^2_{\rm BF}[\Delta f(\xi)] \leq 0.
\end{cases}
\label{eq:minimization}
\ee
In short, we are seeking minimal extensions to the $\Lambda$CDM model as solutions to the Hubble tension which do not deteriorate the fit to a given dataset compared to the $\Lambda$CDM model. While other strategies can be considered (e.g., minimizing the total $\chi^2$ including SH0ES data) as mentioned in Lee \textit{et al.}~\cite{Lee:2022gzh}, we focus on the problem defined in Eq.~\eqref{eq:minimization} to study the existence of scale-dependent modifications to primordial power spectrum which would result in larger inferred value of $H_0$ even without any prior information on $H_0$ preferring its larger value. We define $f$ as the logarithmic dimensionless primordial power spectrum: $f\equiv  \ln \mathcal{P}$. In the standard $\Lambda$CDM model supplemented by SR inflation, it is parameterized as the nearly scale-invariant power spectrum, $\mathcal{P}_{0}(k) = A_s (k/k_p)^{n_s-1}$, with the pivot scale conventionally defined as $k_p\equiv 0.05\;\text{Mpc}^{-1}$. We write $f$ as a function of $\xi \equiv k \eta_0$, where $\eta_0$ is the conformal time today, as we justify later, and $||\Delta f(\xi)||^2 \equiv \int d\ln \xi\; \big[\Delta f(\xi)\big]^2$ is the L2 norm in $\ln \xi$. 

Although the minimization problem given in Eq.~\eqref{eq:minimization} can be solved exactly, it is computationally demanding. To bypass this issue,  Lee \textit{et al.}~\cite{Lee:2022gzh} take a perturbative approach, building on the Fisher-bias formalism: extensions to the $\Lambda$CDM model $\Delta f(\xi)$ are assumed to be small, enabling one to Taylor-expand the effect of perturbations in $f$ on predictions for cosmological observables. Starting from a Gaussian chi-squared with inverse covariance matrix $\bm{M}$, 
\be
\chi^2(\vec{\Omega}) \equiv [\bm{X}(\vec{\Omega}) - \bm{X}^{\rm obs}] \cdot \bm{M}(\vec{\Omega}) \cdot [\bm{X}(\vec{\Omega}) - \bm{X}^{\rm obs}],
\label{eq:chi-square}
\ee
where $\bm{X}(\vec{\Omega})$ is a theory prediction and $\bm{X}^{\rm obs}$ is the data, and expanding to second order in $\Delta f(\xi)$, Lee \textit{et al.}~\cite{Lee:2022gzh} derive the following expressions for shifts in best-fit cosmology away from some fiducial values $\vec{\Omega}_{\rm fid}$, and change in the best-fit chi-square:
\barr
\Delta \Omega_{\rm BF}^i &=& \int d\ln \xi\; \frac{\delta \Omega_{\rm BF}^i}{\delta  f(\xi)} \Delta f(\xi),\label{eq:DObf_X}\\
\Delta \chi_{\rm BF}^2 &=& \int d\ln \xi\; \frac{\delta \chi_{\rm BF}^2}{\delta f(\xi)}\Delta f(\xi)\label{eq:bf} \nonumber\\
&+& \frac12 \iint d\ln \xi\;d\ln \xi'\; \frac{\delta^2 \chi_{\rm BF}^2}{\delta f(\xi) \delta f(\xi')} \Delta f(\xi) \Delta f(\xi'),~~\label{eq:Dchi2bf_X}~~
\earr
where
\barr
\frac{\delta \Omega_{\rm BF}^i}{\delta f(\xi)} &=& -(F^{-1})_{ij} \frac{\partial \bm{X}}{\partial \Omega^j} \cdot \bm{M}\cdot\frac{\delta \bm{X}}{\delta f(\xi)},\label{eq:dObf}\\
\frac{\delta \chi_{\rm BF}^2}{\delta f(\xi)} &=& 2[\bm{X}(\vec{\Omega}_{\rm fid}) - \bm{X}^{\rm obs}] \cdot\widetilde{\bm{M}} \cdot \frac{\delta \bm{X}}{\delta f(\xi)},\label{eq:dchi2bf_lin}\\
\frac{\delta^2 \chi_{\rm BF}^2}{\delta f(\xi) \delta f(\xi')} &=& 2 \frac{\delta \bm{X}}{\delta f(\xi)}\cdot\widetilde{\bm{M}}\cdot\frac{\delta \bm{X}}{\delta f(\xi')},\label{eq:dchi2bf_quad}
\earr
and
\be
\widetilde{M}_{\alpha\beta} \equiv M_{\alpha\beta} - M_{\alpha \gamma} \frac{\partial X^\gamma}{\partial \Omega^i}(F^{-1})_{ij}\frac{\partial X^\sigma}{\partial \Omega^j}M_{\sigma\beta}.
\ee
Note that the subscript BF stands for best-fit. We refer to Lee \textit{et al.}~\cite{Lee:2022gzh} for details about the expressions above.

To go beyond linearity, we apply this method iteratively so that we can obtain a more precise final solution, which is expected to converge to the exact one with a large enough number of iterations.

In the following, we define $\xi\equiv k\eta_0$ where $\eta_0$ is the conformal time today. This is to make the effect of localized changes in the primordial power spectrum on CMB anisotropy spectra, which are the primary data in this study, as independent of each other as possible even when cosmologies are varied. This follows from the fact that the $C_\ell$'s are photon intensity/polarization power spectra projected on 2D spherical surface (or integrated along line-of-sight) involving the spherical Bessel function $j_\ell(k\eta_0)$\footnote{Here, $\eta_0$ is approximated from $\eta_0-\eta_*$ since $\eta_0\gg\eta_*$ where $\eta_*$ is the conformal time at the last scattering surface.}, with cosmology-dependence in $\eta_0$.

Similarly to what is done in Lee \textit{et al.}~\cite{Lee:2022gzh}, we calculate the functional derivatives of $\bm{X}$ using localized Gaussian functions\footnote{Note that one could use other form of functions instead of Gaussian, and we checked that using, for example, trigonometric functions work as well.} for $f(\xi)$ centered on discrete values $\xi_i$. In more detail, we define $N =1,000$ equally-spaced Dirac-delta-like functions in $\log_{10}\xi = \log_{10}(k \eta_0)$, in the range $(\xi_{\min}, \xi_{\max})$ where $\xi_{\min} = (5\times10^{-5}\;\text{Mpc}^{-1})\eta_0^{\rm fid}$ and $\xi_{\max} = (0.5\;\text{Mpc}^{-1})\eta_0^{\rm fid}$ and $\eta_0^{\rm fid}$ is the conformal time today for the fiducial cosmology, which we set to be the Planck $\Lambda$CDM best-fit \cite{Planck2018}:
\barr
f(\xi,\xi_i) \equiv \Delta \ln \mathcal{P}(\xi,\xi_i) \propto \exp{\left[ - \frac{[\log_{10}(\xi/\xi_i)]^2}{2\sigma^2}\right]},
\earr
where $\sigma \equiv [\log_{10}(\xi_{\max}/\xi_{\min})]/N$. Using these functions, we calculate the CMB anisotropy spectra $C_\ell[\vec{\Omega}, \pm\Delta\ln \mathcal{P}(\xi,\xi_i)]$ for each $\xi_i$ and perform the two-sided numerical derivatives. The CMB spectra are computed using a modified version of \textsc{class} \cite{class}, with \textsc{hyrec-2} \cite{Ali-Haimoud:2010tlj, Ali-Haimoud:2010hou, Lee:2020obi} implemented for the recombination history.

\section{Data}
\label{sec:data}
Before we apply our improved method and present the results, we now go on to specify the data considered in this work. \\

{\bf CMB -- Planck-T\&E (P-T\&E).}
We use Planck 2018 binned spectra (\texttt{cl\_cmb\_plik\_v22.dat}) and covariance matrix (\texttt{c\_matrix\_plik\_v22.dat}) \cite{Planck2018} with $\ell_{\rm min}=30$ and $\ell_{\rm max} = 2508$ for temperature and $\ell_{\rm max}=1996$ for polarization. For low-$\ell$ ($\ell < 30$) TT and EE spectra, we adopt the compressed low-$\ell$ Planck likelihood from Ref.~\cite{Prince:2021fdv}\footnote{\href{https://github.com/heatherprince/planck-low-py}{https://github.com/heatherprince/planck-low-py}} in which the likelihood for binned spectra $D_\ell \equiv \ell(\ell+1)C_\ell/2 \pi$ are given as log-normal probability distribution,
\be
\mathcal{L}(x) = p(x) = \frac{1}{(x-x_0) \sigma \sqrt{2\pi}} e^{-[\ln (x-x_0) - \mu]^2/(2\sigma^2)},
\ee
where $x=D_{\rm bin}$ with two and three bins for TT and EE spectrum, respectively. The values of $x_0$, $\mu$, and $\sigma$ are determined in Ref.~\cite{Prince:2021fdv} having the best-fit log-normal distribution. In the following, we write the chi-squared from this likelihood as 
\barr
\chi^2_{\text{low-}\ell} &\equiv& -2\ln \mathcal{L}(x) \nonumber \\
&=& \frac{ [\ln (x-x_0) - \mu + \sigma^2]^2}{\sigma^2} + \text{constant},
\earr
and ignore constant contributions. We denote our baseline dataset combining low-$\ell$ and high-$\ell$ data as ``Planck-T\&E" or ``P-T\&E".

{\bf CMB lensing -- Planck-lensing (P-lensing).}
We use Planck 2018 CMB lensing measurements \cite{Planck:2018lbu} as additional CMB data. The data is given as binned lensing potential power spectrum, which we take from \texttt{Cobaya} \cite{Torrado:2020dgo}\footnote{\href{https://github.com/CobayaSampler/planck_supp_data_and_covmats/tree/master/lensing/2018}{https://github.com/CobayaSampler/planck\_supp\_data\_and\_\\covmats/tree/master/lensing/2018}}.
\\

{\bf BAO -- BOSS DR12.}
As BAO data, we use BOSS DR12 anisotropic measurements,
\be
\Bigg\{\frac{D_M(z_{\rm eff})r_d^{\rm fid}}{r_d},\frac{H(z_{\rm eff}) r_d}{r_d^{\rm fid}}\Bigg\},
\ee
at three effective redshifts $z_{\rm eff}=0.38~,0.51,~0.61$ \cite{BOSS:2016wmc}\footnote{\href{https://data.sdss.org/sas/dr12/boss/papers/clustering/ALAM_ET_AL_2016_consensus_and_individual_Gaussian_constraints.tar.gz}{https://data.sdss.org/sas/dr12/boss/papers/clustering/\\ALAM\_ET\_AL\_2016\_consensus\_and\_individual\_Gaussian\\\_constraints.tar.gz}}. The definition of the sound horizon at baryon drag $r_d$ we use is that of \textsc{class} \cite{class}, which is based on finding the exact point where the integral $\int_0^{\eta_d} d\eta\;a n_e \sigma_T /R$ reaches unity, where $n_e$ is the number density of electrons, $\sigma_T$ is the Thomson cross section, and $R\equiv3\rho_b/4\rho_\gamma$. \\

{\bf Uncalibrated SN\Romannum{1}a -- PantheonPlus.}
We consider the constraint on the energy density of total matter $\Omega_m = 0.334 \pm 0.018$ from PantheonPlus \cite{Brout:2022vxf}. We write
\be
\Omega_m = (\omega_c + \omega_b + \omega_\nu) h^{-2},
\ee
where $\omega_\nu = 0.000644$ is fixed with one massive neutrino species of $m_\nu=0.06$ eV, and include the constraints on it in our dataset.

For completeness, we show all the functional derivatives we calculate using Eq.~\eqref{eq:dObf}--\eqref{eq:dchi2bf_quad} with the data provided above in Appendix.~\ref{appendix:deriv}.

\begin{figure*}[!t]
\centering
\includegraphics[width = \linewidth,trim= 5 5 0 0]{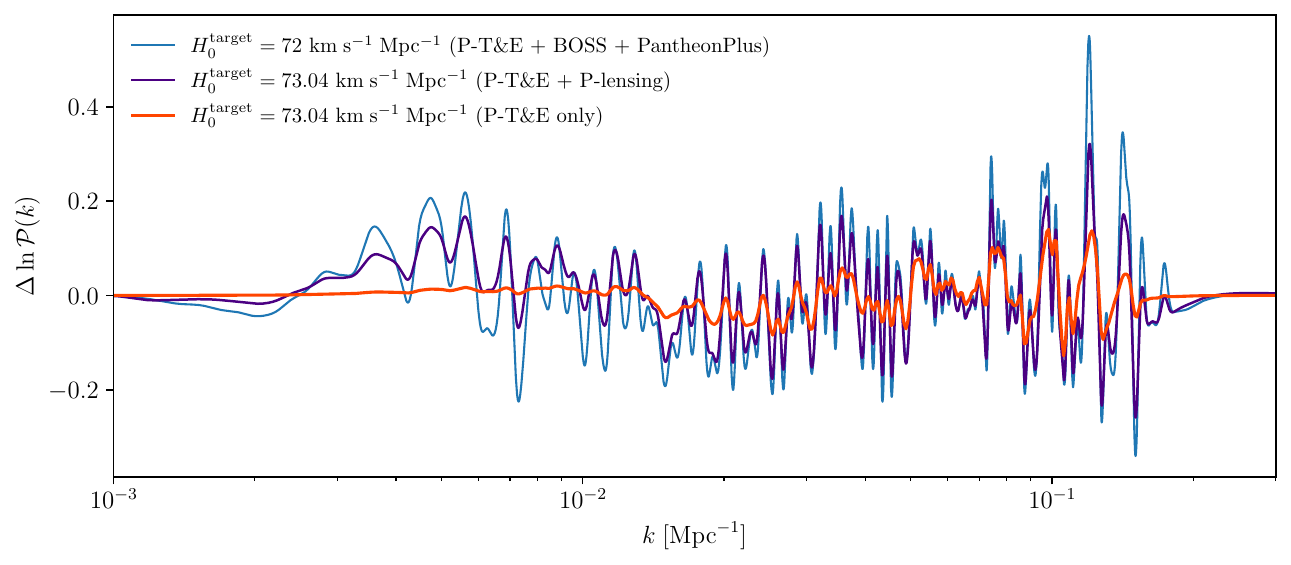}
\includegraphics[width = .49\linewidth,trim= 10 20 0 0]{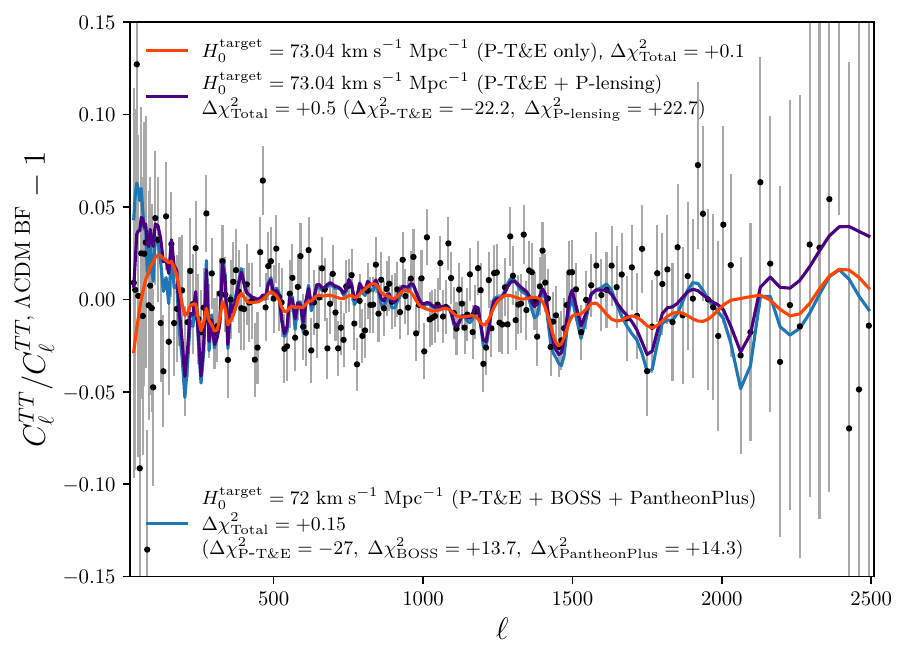}
\includegraphics[width = .48\linewidth,trim= 0 20 10 0]{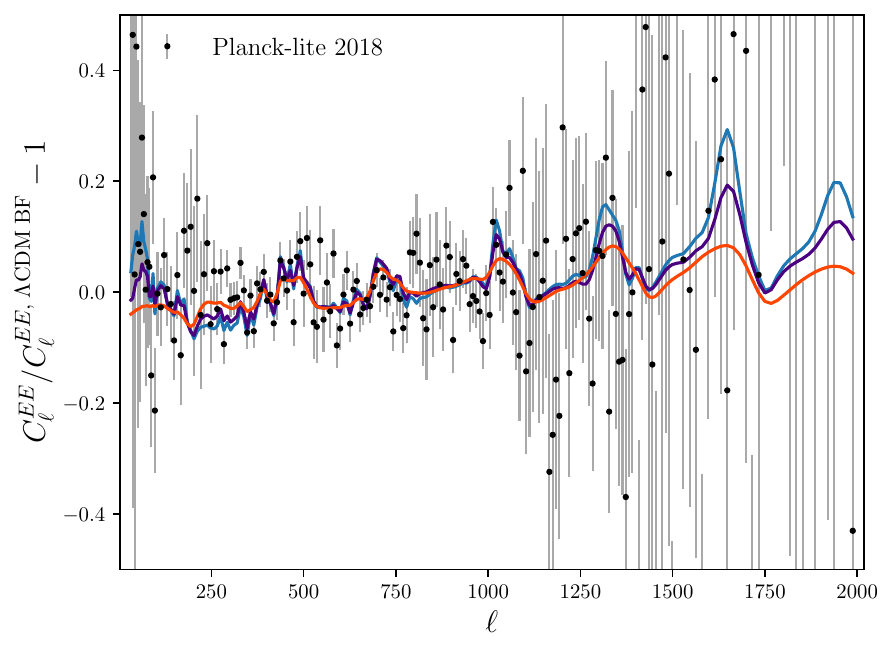}
\caption{Fractional modifications to the primordial power spectrum constructed to result in a best-fit $H_0=73.04\;\text{km\;s}^{-1}\text{Mpc}^{-1}$ with Planck-T\&E data (orange curves) and Planck-T\&E + Planck-lensing data (purple curves), and $H_0=72\;\text{km\;s}^{-1}\text{Mpc}^{-1}$ with Planck-T\&E + BOSS BAO + PantheonPlus SNIa data (blue curves). The fractional residuals of resulting CMB TT/EE spectra with respect to the Planck $\Lambda$CDM best-fit are shown in the bottom panels together with Planck data. The changes in the total best-fit $\chi^2$ are given, which are consistent with the changes estimated by our method, Eq.~\eqref{eq:Dchi2bf_X}, and shown in Table \ref{tab:sol}.}
\label{fig:CMB-CMBBAOPantheon}
\end{figure*}

\begin{table*}[ht!]
  \centering
  \begin{tabular}{c|c|c|c|c|c|c}
   Dataset & \multicolumn{2}{c|} {P-T\&E} & \multicolumn{2}{c|} {~~~P-T\&E + P-lensing~~~}& \multicolumn{2}{c} {~~~P-T\&E + BOSS + PantheonPlus~~~}\\
    \hline
  Model & ~~~~~~~$\Lambda$CDM~~~~~~~ & ~$\Delta \ln \pk$ ~&  ~~~~~~~$\Lambda$CDM~~~~~~~
 & ~$\Delta \ln \pk$~ &  ~~~~~~$\Lambda$CDM~~~~~~
 & $\Delta \ln \pk$ \\
  \hline
  \hline
  $H_0$ (km/s/Mpc)   & $67.20\pm0.66$ &~$73.12\pm0.50$~ & $67.34\pm0.53$ &$73.20\pm0.65$ & $67.31\pm0.43$ &$72.19\pm0.40$\\
    \hline
     $\omega_c$ & $0.1204\pm0.0015$ & $0.1087\pm0.0010$ & $0.1200\pm0.0012$ & $0.1095\pm0.0012$& $0.1200\pm0.0010$ & $0.1120\pm0.0007$ \\
    \hline
     $100\omega_b$ & $2.237\pm0.016$ & $2.380\pm0.014$ & $2.237\pm0.015$ & $2.426\pm0.018$& $2.235\pm0.012$ & $2.426\pm0.014$ \\
    \hline
     $\tau$ & $0.0528\pm0.0075$ & $0.0624\pm0.0090$ & $0.0538\pm0.0075$  & $0.0866\pm0.0109$ & $0.0597\pm0.0073$  & $0.0506\pm0.0087$ \\
    \hline
     $\ln (10^{10}A_s)$ & $3.043\pm0.015$ & $3.033\pm0.017$ &  $3.044\pm0.014$ & $3.101\pm0.021$&  $3.055\pm0.015$ & $3.022\pm0.018$\\
    \hline
     $n_s$ & $0.9651\pm0.0047$ & $0.9826\pm0.0036$ & $0.9654\pm0.0042$  & $0.9963\pm0.0048$& $0.9657\pm0.0040$  & $0.9670\pm0.0039$\\
  \end{tabular}
  \caption{Constraints on cosmological parameters for three different sets of data, under two different models. Here Planck-T\&E, Planck-lensing, BOSS, PantheonPlus refer to Planck DR3 lite-likelihood together with the compressed low-$\ell$ TT and EE likelihood from Ref.~\cite{Prince:2021fdv}, Planck lensing likelihood \cite{Planck:2018lbu}, BOSS DR12 anisotropic measurements \cite{BOSS:2016wmc}, and PantheonPlus uncalibrated SN\Romannum{1}a \cite{Brout:2022vxf}, respectively. The first columns under each set of data present the constraints under $\Lambda$CDM, and the second columns show the constraints with our solution obtained given each dataset (which are shown in the top panel of Fig.~\ref{fig:CMB-CMBBAOPantheon}).}
\label{tab:sol}
\end{table*}

\section{Results}
\label{sec:results}

\subsection{Application to Planck CMB data}

\subsubsection{Planck-T\&E (P-T\&E)}
\label{sec:results-Planck-lite}

We solve the optimization problem we set in Eq.~\eqref{eq:minimization} given the target values for the new best-fit Hubble constant $H_0^{\rm target}$. While it can be solved exactly if combined with Markov Chain Monte Carlo (MCMC) analysis exploring arbitrary $\Delta f$, this is computationally expensive. We instead take the Fisher-bias formalism described in Sec.~\ref{sec:method}, and apply it iteratively. This iterative method becomes equivalent to exactly solving the given optimization problem for a large enough number of iterations, but in practice we found that at most two iterations are enough. Specifically, when $H_0^{\rm target}>70.15\;\text{km\;s}^{-1}\text{Mpc}^{-1}$---i.e. the halfway value from Planck best-fit to SH0ES best-fit---we first construct a solution with the target Hubble constant $H_0^{\rm target}=70.15\;\text{km\;s}^{-1}\text{Mpc}^{-1}$, we set the new best-fit cosmology under this solution as our new fiducial cosmology, and then we construct another solution on top of the first one to finally achieve $H_0^{\rm \rm target}>70.15\;\text{km\;s}^{-1}\text{Mpc}^{-1}$. 

We first focus on finding a particular solution which accommodates the SH0ES best-fit as our target value of $H_0^{\rm target}=73.04\;\text{km\;s}^{-1}\text{Mpc}^{-1}$. This solution, which entirely resolves the Hubble tension between Planck-T\&E and SH0ES, is shown as the orange line in the upper panel of Fig.~\ref{fig:CMB-CMBBAOPantheon}. Although the shape of this modification to $\mathcal{P}(k)$ is non-trivial, this solution still results in relatively smooth changes in CMB spectra with accordingly shifted cosmological parameters---see orange lines in the bottom panels of Fig.~\ref{fig:CMB-CMBBAOPantheon}. That is, while the underlying theoretical explanation still needs to be properly modeled, it implies that non-trivial modifications to primordial power spectrum have the potential to pin down a very different best-fit cosmology compared to that with the standard nearly scale-invariant power spectrum, yet providing theoretical predictions of CMB spectra consistent with Planck measurements.

\begin{figure*}[ht!]
\centering
\includegraphics[width = .9\linewidth,trim= 10 20 10 10]{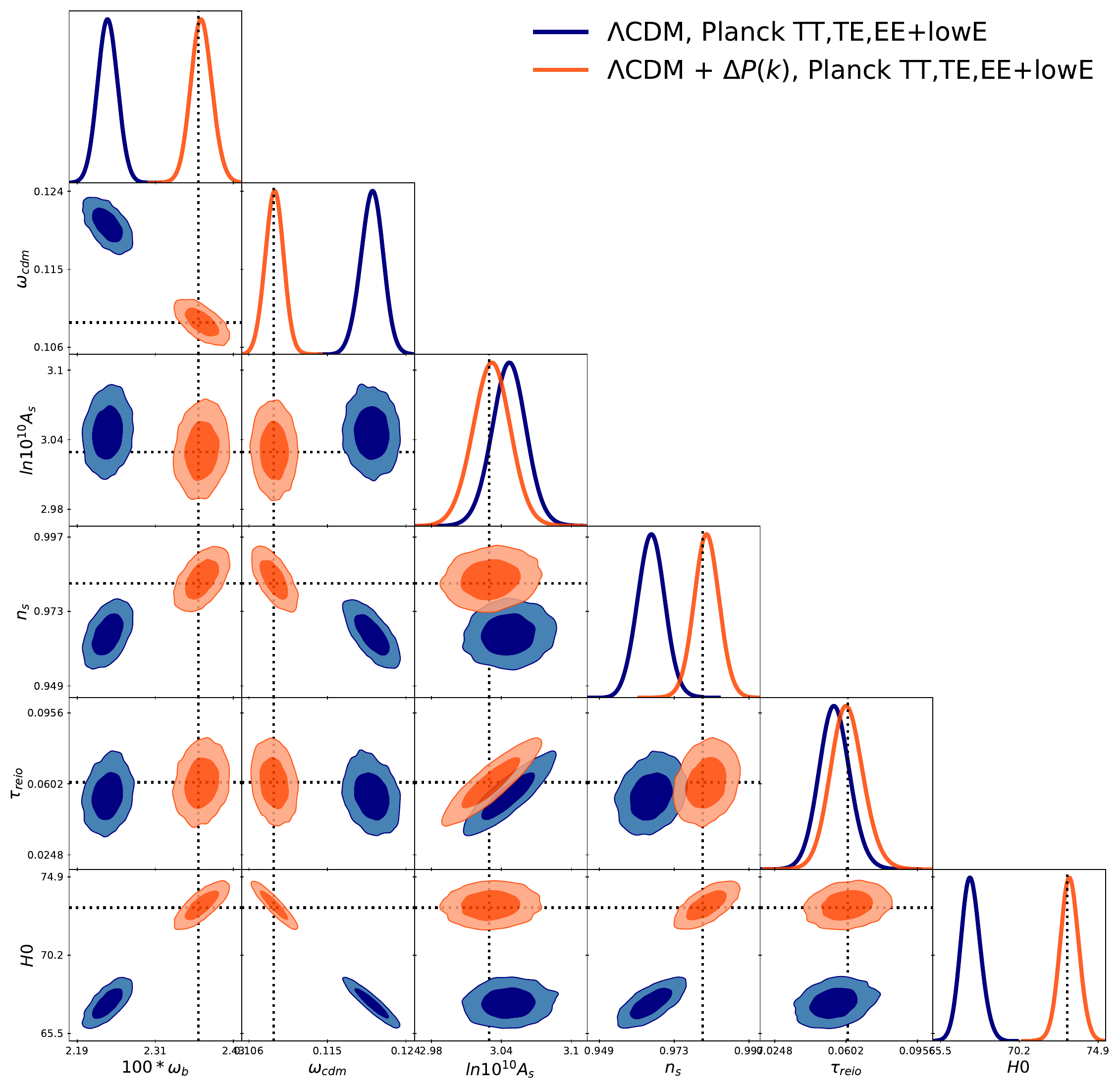}
\caption{Contour plot for $\Lambda$CDM and $\Lambda$CDM + $\Delta \ln \mathcal{P}(k)$. In the latter model, we fix $\Delta \ln \mathcal{P}(k)$ to the orange line in the top left panel of Fig.~\ref{fig:CMB-CMBBAOPantheon}, which solves the tension between  SH0ES and Planck-T\&E data. The black dotted lines are the estimated new best-fits by our formalism, Eq.~\eqref{eq:bf}, which well agree with the peaks of 1D posteriors from MCMC.}
\label{fig:MCMC}
\end{figure*}

\begin{figure}[t!]
\includegraphics[width = .95\columnwidth,trim= 20 20 20 0]{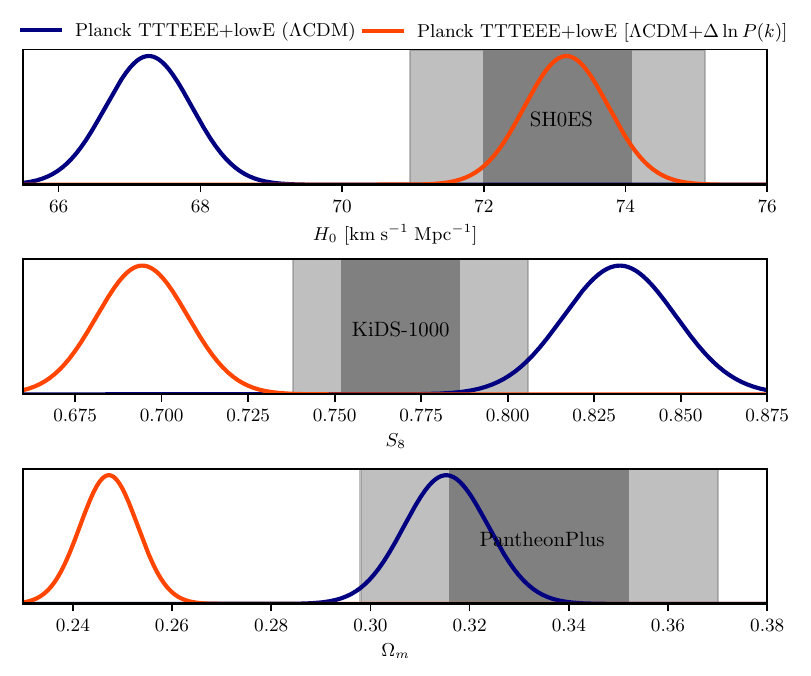}
\caption{Posteriors of $H_0$ (top), $S_8$ (middle), and $\Omega_m$ (bottom) inferred from Planck TTTEEE + lowE likelihood, together with SH0ES, KiDS-1000, and PantheonPlus results, shown as grey bands.}
\label{fig:H0Om}
\end{figure}

Our solution is found by solving an optimization problem using the Fisher-bias formalism, with an iterative approach. While this method is expected to be equivalent to solving such an optimization problem exactly, it is still an approximate method with a finite number of iterations. Hence, to confirm the validity of our solution, we perform an MCMC analysis using \textsc{montepython} v3.0 \cite{Audren:2012wb,Brinckmann:2018cvx} and show the results in Fig.~\ref{fig:MCMC} and Table.~\ref{tab:sol}. As shown in Fig.~\ref{fig:MCMC}, where the estimated new best-fit is shown as dotted lines, the solution shifts the best-fit cosmological parameters as estimated by the Fisher-bias formalism [Eq.~\eqref{eq:Dchi2bf_X}], demonstrating the validity of our method. The posterior of $H_0$ obtained from the MCMC analysis with this solution, which is shown in the top panel of Fig.~\ref{fig:H0Om}, implies that the tension is indeed fully resolved. 

So far, we have simply addressed the question of whether CMB data can accommodate a larger best-fit value of $H_0$ by allowing the primordial power spectrum to have spectral features beyond near scale-invariance. The answer to this question is {\em yes}, suggesting that the tension between SH0ES and Planck data can in principle be resolved by non-standard primordial physics. This is a non-trivial result: it could very well have been the case that no modification of $\mathcal{P}(k)$ can ever mimic an increase in $H_0$ as inferred from the CMB. However, this increased $H_0$ requires significant shifts in some of the other cosmological parameters, which may then bring them in tension with measurements from other datasets, thus making the ``solution" of a modified $\mathcal{P}(k)$ less appealing.

First, we find that the cosmological model best fitting CMB data with the perturbed $\mathcal{P}(k)$ has a significantly decreased value of $\Omega_m=0.247$, producing a new tension with the PantheonPlus SNIa measurement (see the bottom panel of Fig.~\ref{fig:H0Om}). The tendency of having lower $\Omega_m$ with modifications in primordial power spectrum resembles the case of modifications of recombination history through a time-varying electron mass and fine structure constant in Lee \textit{et al.}~\cite{Lee:2022gzh} (see also Ref.~\cite{Lynch:2024gmp} in which the same degeneracy is seen with their phenomenological model directly varying recombination history), although a much lower $\omega_c$ (see Fig.~\ref{fig:MCMC}) in our solution results in a correspondingly lower $\Omega_m$. It would be worth noting that this situation is opposite to other proposed solutions to the $H_0$ tension such as EDE, early modifications to gravity or extra relativistic degrees of freedom, where $\Omega_m$ increases while $H_0$ increases~\cite{Schoneberg:2021qvd}. This can be explained with the known dependence of the CMB sound horizon angular scale $\theta_s$ on $\Omega_m$, $h$, and the redshift at recombination $z_*$: $\theta_s\propto (\Omega_m h^3 z_*^{-3.88})^{0.17}$. While this dependence is determined under $\Lambda$CDM model, it is still valid in our case as all we modify is the primordial power spectrum leaving the degeneracies among such parameters untouched. Lowering $\Omega_m$ while increasing $h$ is thus inevitable since our model does not affect the recombination process.

As a consequence of the lowered $\Omega_m$, the parameter $S_8 = \sigma_8 (\Omega_m/0.3)^{0.5}$ is also significantly lowered, where $\sigma_8$ is the amplitude of matter clustering at the scale of 8 Mpc/$h$. While in the case of perturbed recombination \cite{Lee:2022gzh}, this helped alleviate the well-known $S_8$ tension between weak lensing probes and Planck CMB anisotropy data \cite{DES:2017myr,KiDS:2020suj,DiValentino:2020vvd}, in the present case, the CMB best-fit $S_8$ is lowered so much that it now falls significantly below weak-lensing measurements, to the point that the $S_8$ tension is reversed in sign, and slightly worsened, as shown in the middle panel of Fig.~\ref{fig:H0Om}.

Lastly, the modification of $\mathcal{P}(k)$ we found implies a significantly increased baryon density $\omega_b=0.02380\pm0.00014$ leading to a $2.8\sigma$ discrepancy with the recently updated BBN constraint $\omega_b=0.02218 \pm 0.00055$ \cite{Schoneberg:2024ifp}.

Such primordial features could be constrained by large-scale structure measurements as well \cite{Beutler:2019ojk,Calderon:2025xod}, but the main constraining factor would be the shift of best-fit cosmology under CMB data, rather than the primordial features itself.

\subsubsection{Planck-T\&E (P-T\&E) + Planck-lensing (P-lensing)}
\label{sec:results-Planck-lite-lensing}
\begin{figure}[t!]
\includegraphics[width = .95\columnwidth,trim= 20 20 20 0]{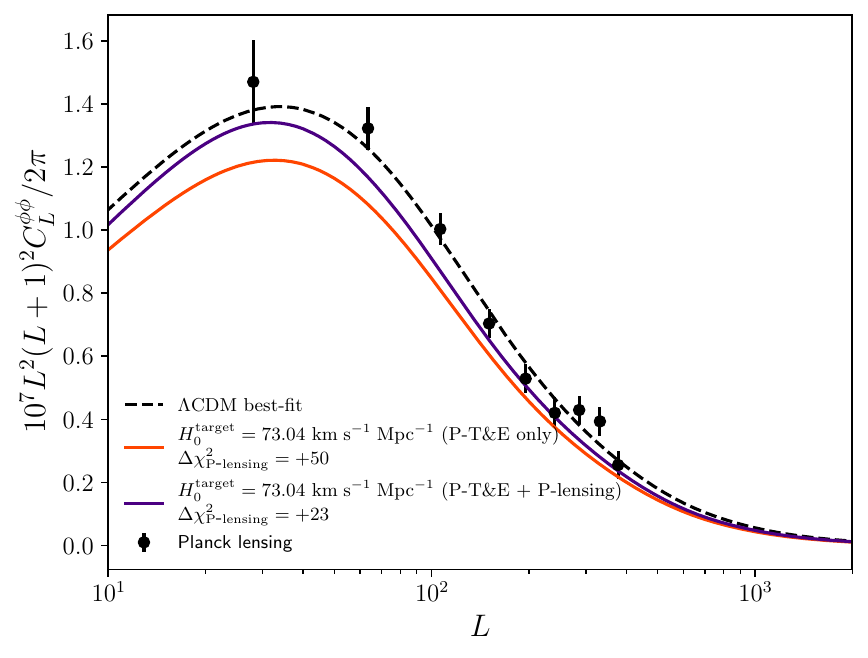}
\caption{CMB lensing-potential power spectrum. Our Planck-T\&E only solution (orange) provides a worse fit to Planck CMB lensing measurements ($\Delta\chi^2_{\rm P\textrm{-}lensing}=+50$) compared to $\Lambda$CDM best-fit. While the solution with Planck-T\&E + Planck-lensing data keeps the total chi-squared effectively unchanged, it still results in a worse fit ($\Delta\chi^2_{\rm P\textrm{-}lensing}=+23$).}
\label{fig:phiphi}
\end{figure}

The lower value of $\Omega_m$ from the Planck-T\&E solution in Sec.~\ref{sec:results-Planck-lite} potentially worsens the fit to datasets such as Planck lensing, BOSS, and PantheonPlus, all of which favor higher values of $\Omega_m$. In particular, Fig.~\ref{fig:phiphi} shows that our best-fit solution degrades the fit to Planck lensing data by $\Delta \chi^2_{\rm P\textrm{-}lensing} = +50$. In this subsection, we therefore investigate whether an alternative solution can better accommodate lensing data. To that end, we include Planck lensing measurements as described in Sec.~\ref{sec:data}. We find that it is still possible to achieve $H_0 = 73.04\;\text{km\,s}^{-1}\text{Mpc}^{-1}$ while keeping the total chi-squared unchanged. However, this comes at the cost of a significantly worse fit to the lensing measurements ($\Delta \chi^2_{\rm P\textrm{-}lensing} = +23$). This implies that modifications to the primordial power spectrum that raise the CMB-inferred value of $H_0$ tend to be less consistent with Planck lensing and require overfitting Planck-T\&E data to compensate. Interestingly, this outcome is not necessarily expected a priori, since lensing measurements are largely insensitive to oscillations in the primordial power spectrum, which are smoothed out due to the broad range of contributing wave numbers at a given multipole $\ell$, see e.g.~\cite{Chluba:2015bqa}. However, the modified solution induces strong anti-correlations between $\Omega_m$ and $H_0$, which do affect the lensing potential power spectrum—highlighting the importance of including CMB lensing in such analyses.

\subsection{Application to Planck CMB + BAO + PantheonPlus}

Given that the solution constructed to resolve the Hubble tension using only Planck-T\&E data does not spontaneously encompass other cosmological data that prefer higher $\Omega_m$, we include BOSS BAO and PantheonPlus SNIa data in our data vector $\bm{X}$ and repeat the analysis to see if we can construct solutions to resolve the Hubble tension that maintain consistency among these data as in the standard $\Lambda$CDM model. With these data included, we could achieve up to $H_0=72\;\text{km\;s}^{-1}\text{Mpc}^{-1}$ as our minimizer fails to find converging solutions to our optimization problem in Eq.~\eqref{eq:minimization} with higher target values of $H_0$.  This is similar to our findings in Sec.~\ref{sec:results-Planck-lite-lensing}. However, the dataset considered here imposes tighter constraints on $\Omega_m$ than CMB lensing, leading to a stronger enhancement in both  the amplitude and frequency of the oscillations in $\mathcal{P}(k)$  as we target higher values of $H_0$. As a result, our minimizer fails to find solutions.
This is partly related to the degeneracy between $H_0$ and $\Omega_m$ we saw in the previous section. As lowering $\Omega_m$ is inevitable to increase $H_0$, the solution tends to overfit the CMB data to compensate the increase in $\chi^2$ due to lower $\Omega_m$ making the solution further nontrivial. As an example, the solution with this highest achievable target value is shown in the upper panel of Fig.~\ref{fig:CMB-CMBBAOPantheon} as a blue curve, and the resulting differences in the CMB spectra with respect to those of the Planck $\Lambda$CDM best-fit cosmology are given in the bottom panels with the same color. While the resulting change in the total chi-squared $\Delta \chi^2_{\rm Total}=+0.15$ is small, the individual chi-squared for each data ($\Delta \chi^2_{\rm P\textrm{-}T\&E} = -27$, $\Delta\chi^2_{\rm BOSS}=+13.7$, $\Delta \chi^2_{\rm PantheonPlus}=+14.3$) indeed change quite a lot, over-fitting CMB data to compensate for the worsened fit to BAO and SNIa data due to the lowered value of $\Omega_m$. We thus conclude that no modification of the primordial power spectrum can entirely resolve the Hubble tension between Planck CMB and SH0ES, while simultaneously being consistent with other current cosmological data sets which put additional constraints on $\Omega_m$, such as BAO and uncalibrated SN\Romannum{1}a data.

We checked that increasing the number of iterations in our method did not improve the situation, implying that this failure of finding solutions with BAO and SN\Romannum{1}a data included is not due to the approximations we made in the method we take from Lee \textit{et al.}~\cite{Lee:2022gzh}, such as Taylor-expanding $\chi^2$ and the linearity of $\Delta \ln \mathcal{P}$. Furthermore, increasing $N$, the number of $\xi\equiv k\eta_0$ points at which we modify $\mathcal{P}(\xi)$, will not provide additional flexibility to achieve larger $H_0$, as we have a finite number of data points and $N=1,000$ already saturates the flexibility provided by Planck-lite (binned spectra) data.

\section{Implications for inflation}
\label{sec:implications}

Despite the fact that the model does not seem to provide a compelling solution to the Hubble tension when datasets other than Planck and SH0ES are taken into account, it does prove capable of at least increasing $H_0$ compared to the estimate from CMB data under $\Lambda$CDM model. Since the debate regarding systematic effects is far from settled, the model could regain relevance as a solution if, for example, future measurements by SH0ES were to shift towards smaller values of $H_0$. For this reason, we find it both worthwhile and instructive to explore what kinds of mechanisms could lead to the modification of the primordial power spectrum presented above. In this Section, we discuss the implications of our findings on the primordial power spectrum $\mathcal{P}(k)$ for inflationary model building. In general, oscillatory features can arise from two main mechanisms, as extensively discussed in the literature (see e.g. the reviews~\cite{Chen:2010xka,Chluba:2015bqa,Slosar:2019gvt,Achucarro:2022qrl}). 

The first mechanism involves a temporary departure of a background quantity, denoted as $ B(t)$, from its SR attractor during inflation, characterized by $ \dot{B}/HB \ll \mathcal{O}(1)$ for a brief period, typically shorter than 1 $ e $-fold. This departure results in an oscillatory correction of the form $ \Delta\ln \pk \sim A(k) \sin(2k/k_0) $, where $ A(k) $ represents a model-dependent envelope, and $ k_0 $ is the frequency associated with the scale that crosses the Hubble radius at the time of the sharp feature in the background.

\begin{figure*}[ht!]
	\centering
	\includegraphics[width=2\columnwidth]{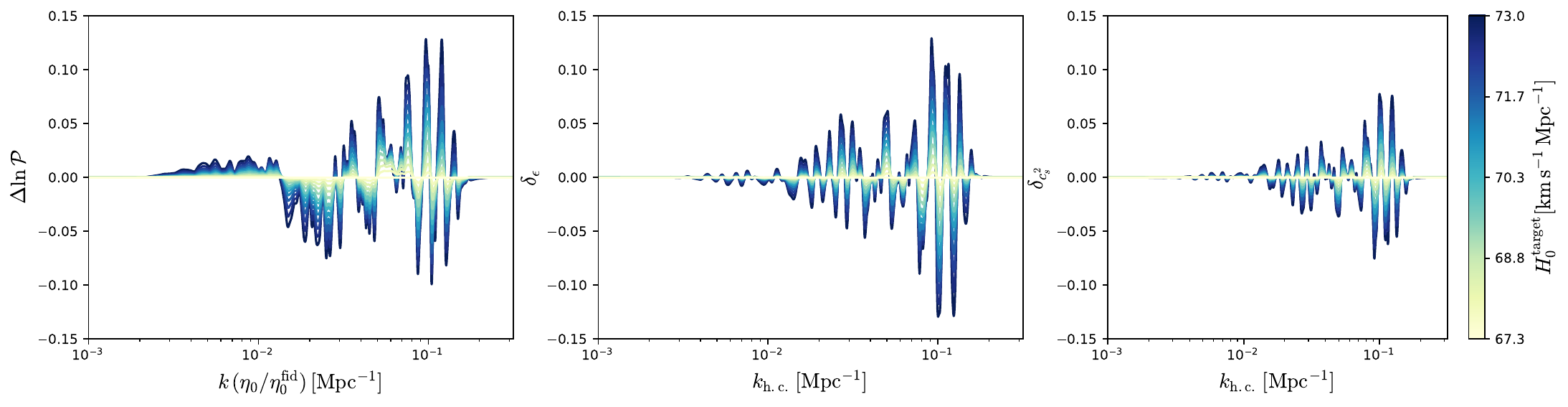}
	\caption{Left: Solutions for $\Delta \ln \pk \equiv \frac{\Delta \mathcal{P}}{\mathcal{P}}(k)$ given target values of the CMB-only best-fit Hubble constant $H_0$, using Planck anisotropy data \cite{Planck2018}. All solutions are constructed to keep the Planck best-fit chi-squared unaffected. Center (Right): Variations of the first SR parameter (speed of sound squared of curvature perturbations) computed from the solutions in the left panel. The scale $k_{\rm h.c.}$ is related to the time of horizon crossing $t_{\rm h.c.}$ by $k_{\rm h.c.}=1/a(t_{\rm h.c.}) H(t_{\rm h.c.})$. }
	\label{fig:sol}
\end{figure*}

The second mechanism involves a small oscillatory correction to a background quantity, where sub-horizon modes resonate with the background when their physical frequency becomes comparable to the oscillation frequency, leading to a correction of the form $ \Delta\ln \pk \sim \sin(\omega/H \ln k/k_*) $, where $ \omega/H $ is the background frequency normalized to the Hubble scale during inflation. Realistic scenarios may exhibit a combination of these two types of signals. For instance, models known as ``primordial standard clocks" manifest as a superposition of sharp and resonant feature signals~\cite{Chen:2014cwa,Braglia:2021ckn}. Additionally, alternative primordial universe scenarios may introduce different runnings of the oscillations~\cite{Chen:2011zf,Chen:2018cgg,Domenech:2020qay,Quintin:2024boj}.

Given that the oscillatory signal in the solutions shown in the left panel of Fig.~\ref{fig:sol} does not distinctly indicate either a sharp or a resonant feature signal, modeling it in terms of these mechanisms may not be straightforward. Instead, we opt for an alternative approach to discuss its relation to inflationary mechanisms. Our approach involves reconstructing the time dependence of the background quantities that could give rise to such a spectrum. This method, previously employed in the literature~\cite{Canas-Herrera:2020mme,Antony:2022ert}, offers the advantage of being more generic and applicable to inflation models that can be effectively described by a single field. 

In particular, we can work in the flexible framework of the EFT of inflationary fluctuations \cite{Creminelli:2006xe,Cheung:2007st}, where we do not need to model the background evolution, and we can simply assume a de Sitter expansion for the inflationary era.  Small scale-dependent corrections in the primordial power spectrum, such as those under consideration, stem from small time-dependence in the coefficients of the quadratic self-interactions of the curvature perturbation. The quadratic Hamiltonian is given by~\cite{Chen:2010xka,Chluba:2015bqa}:
\begin{equation}
	H^{(2)}=M_{\rm pl}^2 a^3 \epsilon\left[\frac{\dot{\zeta}^2}{c_s^2}+(\partial\zeta)^2\right].
\end{equation}
In this equation, $a$ represents the scale factor, $H$ is the Hubble rate during inflation, $\epsilon\equiv-\dot{H}/H^2$ denotes the first SR parameter, and $c_s^2$ is the sound speed of curvature perturbations, which, in general single-field models of inflation, need not be equal to $1$. 

Under the assumption of near de Sitter evolution with unity sound speed, we have $a\simeq-1/H\eta$,  and  $H\sim {\rm constant}$, $\epsilon\equiv \epsilon_0\sim {\rm constant}$ and $c_s^2=1$.

To compute the effect of small deviations from $\epsilon_0$ and $c_s=1$, we split the quadratic Hamiltonian into a {\em free} and {\em interaction} part as follows:
\begin{equation}
H^{(2)}\equiv H^{(2)}_0+\Delta_{\epsilon} H^{(2)}+\Delta_{c_s^2} H^{(2)},
\end{equation}
where  the free Hamiltonian is 
\begin{equation}
	H^{(2)}_0\equiv M_{\rm pl}^2 a^3 \epsilon_0\left[\dot{\zeta}^2+(\partial\zeta)^2\right],
\end{equation}
with $\epsilon_0={\rm constant}$ and we treat remaining quadratic interactions as perturbations
\begin{align}
\Delta_{\epsilon} H^{(2)}\equiv&M_{\rm pl}^2 a^3 \epsilon_0\delta_\epsilon \left[\dot{\zeta}^2+(\partial\zeta)^2\right],\\
\Delta_{c_s^2} H^{(2)}\equiv& M_{\rm pl}^2 a^3 \epsilon_0 \left(\frac{1}{c_s^2}-1\right)\dot{\zeta}^2\equiv M_{\rm pl}^2 a^3 \epsilon_0 \delta_{c_s^2}\dot{\zeta}^2.
\end{align}

The correction to the nearly scale-invariant power spectrum can then be computed perturbatively using the in-in formalism~\cite{Maldacena:2002vr,Weinberg:2005vy,Chen:2010xka}. The leading order correction to the primordial power spectrum is given by\footnote{Higher order corrections, induced by the insertion of $n\geq2$ quadratic vertices in the perturbative expansion for the tree level power spectrum are subdominant for correction to the power spectrum of order $\Delta\ln\pk\sim\mathcal{O}(0.1)$~\cite{Inomata:2022yte}.}~\cite{Braglia:2022ftm}:
\begin{equation}
	\label{eq:delta_eps_cs}
	\Delta_i\ln\pk=\int_0^\infty\dd  k_1 \,\delta_i(k_1) f_i(k_1,\,k)
\end{equation}
where
\barr
	f_\epsilon(k_1,\,k)&=&-\frac{1}{4 k}\Biggl[2\frac{k}{k_1}\cos\left(-2\frac{k}{k_1}\right)\notag\\&&+\left(1-2\frac{k^2}{k_1^2}\right)\sin\left(-2\frac{k}{k_1}\right)\Biggr],\\	
	f_{c_s^2}(k_1,\,k)&=&-\frac{1}{4 k}\frac{k^2}{k_1^2}\sin\left(-2\frac{k}{k_1}\right).
\earr
We can then define 	$x\equiv\ln k$ and $y\equiv \ln k_1$
to recast Eq.~\eqref{eq:delta_eps_cs} into the following form:
\begin{equation}
	\label{eq:delta_eps_cs_fin}
	\Delta_i\ln\mathcal{P}(x)=\int_{-\infty}^\infty\dd  y \,\delta_i(y) g_i(x-y)
\end{equation}

where
\begin{align}
	g_\epsilon(z)&=-\frac{e^{-z}}{4 }\Biggl[2e^{z}\cos\left(-2e^{z}\right)+\left(1-2e^{2 z}\right)\sin\left(-2e^{z}\right)\Biggr]\\	
	g_{c_s^2}(z)&=-\frac{1}{4 }e^{z}\sin\left(-2e^{z}\right).
\end{align}

The advantage of recasting Eq.~\eqref{eq:delta_eps_cs} into~\eqref{eq:delta_eps_cs_fin}, is that  the right hand side can now be written as a convolution of the functions $\delta_i$ and $g_i$ so that the solution for the former function is just given by
\begin{equation}
	\label{eq:solution_perturbation}
	\delta_i(x)=\mathcal{F}_x^{-1}\left[\frac{\mathcal{F}_\omega[\Delta_i\ln\mathcal{P}(x)]}{\mathcal{F}_\omega[g_i(x)]}\right],
\end{equation}
where $\mathcal{F}_\omega$ and $\mathcal{F}_x^{-1} $ denote the Fourier transform and its inverse respectively, and $\omega$ is the variable conjugate to $x$.

The reconstructed background functions are depicted in the middle and right panels of Fig.~\ref{fig:sol}. While their behavior may not be easily described by simple elementary functions or monochromatic oscillations, they offer valuable insights into the inflationary mechanisms at play. Notably, while the last three oscillations at $ k\geq0.1\,{\rm Mpc}^{-1}$ suggest a resonant nature--- fitting well with a background oscillation frequency of $ \omega/H\sim34 $---the rest of the signal, particularly at smaller wavenumbers, cannot be well described by a resonant feature. Since these significantly contribute to the total $\chi^2$---see Fig.~\ref{fig:deriv-chi2} in Appendix \ref{appendix:deriv}---it would be misleading to model the feature as a pure logarithmic oscillation. At these scales, the oscillatory behavior is less clear, possibly indicating interference between the linear oscillations produced by subsequent sharp features observed in our solutions for $\delta_\epsilon $ and $ \delta_{c_s^2} $. While this connection goes beyond the scope of our work, we note that similar features have been proposed in previous studies \cite{Antony:2021bgp,Hazra:2022rdl,Antony:2022ert}.

It is important to note that we have focused primarily on single-field inflationary scenarios, where a direct relation between the power spectrum and a background feature can be established using Eq.~\eqref{eq:solution_perturbation}. More complex multi-field scenarios, such as those involving ``standard clocks"~\cite{Chen:2014cwa} or multiple turns in the field space~\cite{Gao:2015aba}, present challenges in terms of reconstructing the connection between the power spectrum and the background evolution, as the models are specified by more independent background functions, or equivalently of the Wilson-like coefficient of the multifield inflationary EFT~\cite{Pinol:2024arz}. We also note that very sharp, transient violations of slow-roll, such as those generating our power spectrum solutions, may challenge the validity of the EFT description~\cite{Bartolo:2013exa,Cannone:2014qna,Adshead:2014sga,CarrilloGonzalez:2025fqq}. Determining whether this is the case for our scenario, however, requires a dedicated analysis, which lies beyond the scope of this paper.

Finally, let us stress that we have minimized the modifications of primordial power spectrum, $\Delta \ln \mathcal{P}(k)$, in Eq.~\eqref{eq:minimization} as this is the phenomenologically relevant quantity, and we aim to keep the discussion as model-independent as possible with respect to the underlying primordial mechanism. However, one could instead choose to minimize the physical parameters $\delta\epsilon$ and $\delta c_s^2$ directly.

\section{Discussion}
\label{sec:discussion}

In this work, we explored whether a modification of the primordial power spectrum could resolve the Hubble tension by shifting the inferred Hubble constant from Planck CMB data toward the SH0ES measurement, by applying the method developed by Lee \textit{et al.}~\cite{Lee:2022gzh}, further advancing it with an iterative approach.

For the first time, we identified a possible modification that reconciles the Planck-inferred $H_0$ with its direct measurement from SH0ES. However, this solution also lowers the total matter density $\Omega_m$, which is tightly constrained by the BOSS BAO and PantheonPlus uncalibrated SNIa data. Upon incorporating these data sets, we find that achieving the SH0ES-preferred Hubble constant becomes significantly more difficult. One contributing factor is the limited number of degrees of freedom available when fitting the Planck CMB data, restricting the flexibility of modifications to the primordial power spectrum. Additionally, the modifications we have identified tend to overfit individual data points from Planck to compensate for the poor fit to BAO and uncalibrated SNIa when these data are included, further indicating that this approach is not a promising resolution to the Hubble tension. 

We have separately studied how the inclusion of the Planck CMB lensing measurements affects our construction of data-driven solutions to the Hubble tension. While the lensing potential is not sensitive to features in $\pk$ \cite{Chluba:2015bqa} we find that the inclusion of CMB lensing can be important when exploring modifications of $\pk$ as a solution to the Hubble tension, because the degeneracy direction in $h$-$\omega_c$ from the lensing potential is opposite to the shifts in $h$ and $\omega_c$ from the modifications of $\pk$ we found.

We have shown how to translate our results on the primordial power spectrum of curvature fluctuations into insights on the model of the primordial Universe that could have originated them. As the most compelling proposal to date is that of an early inflationary phase, we have worked within the EFT of inflationary fluctuations and showed how our solutions can be produced by a burst of oscillations in the time evolution of either the first slow roll parameter $\epsilon$ or the speed of sound of the curvature perturbation $c_s$.

Our method of identifying modifications to the primordial power spectrum is inherently data-driven, as outlined in Lee \textit{et al.}~\cite{Lee:2022gzh}. The ability to find a solution depends on the errors in the given data, which means that as the precision of measurement improves, the viability of such solutions could change. This underscores the importance of future high-precision large-scale structure surveys and next-generation CMB experiments in refining our understanding of whether modifications to the primordial power spectrum can provide a consistent resolution to the Hubble tension. For example, there have been recent data releases from the Dark Energy Spectroscopic Instrument (DESI) DR1 \& DR2 \cite{DESI:2024mwx,DESI:2024uvr,DESI:2024lzq,DESI:2025zpo,DESI:2025zgx} and Atacama Cosmology Telescope (ACT) DR6 \cite{ACT:2025fju,ACT:2025tim}. While replacing the BOSS BAO data with DESI DR2 would not qualitatively affect our main conclusions—since DESI BAO measurements are largely consistent with BOSS BAO\footnote{Minor impacts could arise. For example, DESI prefers a slightly lower (higher) value of $\Omega_m$ ($H_0$), potentially allowing a modestly higher $H_0$. However, the tighter DESI constraints on $\Omega_m$ may simultaneously make it harder to achieve a large $H_0$.}—the ACT CMB data, which probe smaller angular scales and improve precision at intermediate multipoles, could more strongly constrain modifications to the primordial power spectrum. We leave the investigation with those newly released data for future work. 

\section*{Acknowledgements}

We thank Nils Sch\"oneberg for useful conversations. N.\;L.\;was supported by the James Arthur Graduate Associate Fellowship from the Center for Cosmology and Particle Physics at New York University and the Horizon Fellowship from Johns Hopkins University. This work was supported in part through NYU IT High Performance Computing resources, services, and staff expertise.

\begin{appendix}
\section{Functional derivatives of best-fit parameters and best-fit chi-squared}
\label{appendix:deriv}

\begin{figure*}[ht!]
\centering
\includegraphics[width = .9\linewidth,trim= 00 10 00 20]{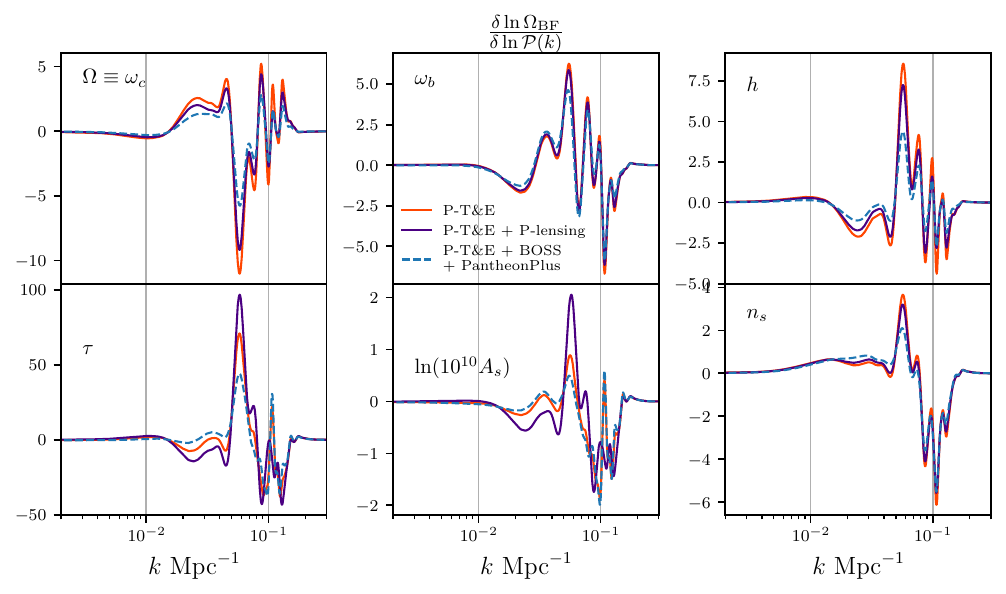}
\caption{Functional derivatives of best-fit parameters, Eq.~\eqref{eq:dObf}, given Planck-T\&E data (orange solid), when either Planck lensing (indigo dot-dashed) or BOSS/PantheonPlus data (blue dashed) is included. We plot with narrower range of $k$ for better presentation  although the range of $k$ we consider for perturbations in $\pk$ is $k \in [5\times 10^{-5},0.5]\;\text{Mpc}^{-1}$.}
\label{fig:deriv-Omega}
\end{figure*}

\begin{figure*}[ht!]
\centering
\includegraphics[width = .33\linewidth,trim= 10 0 00 20]{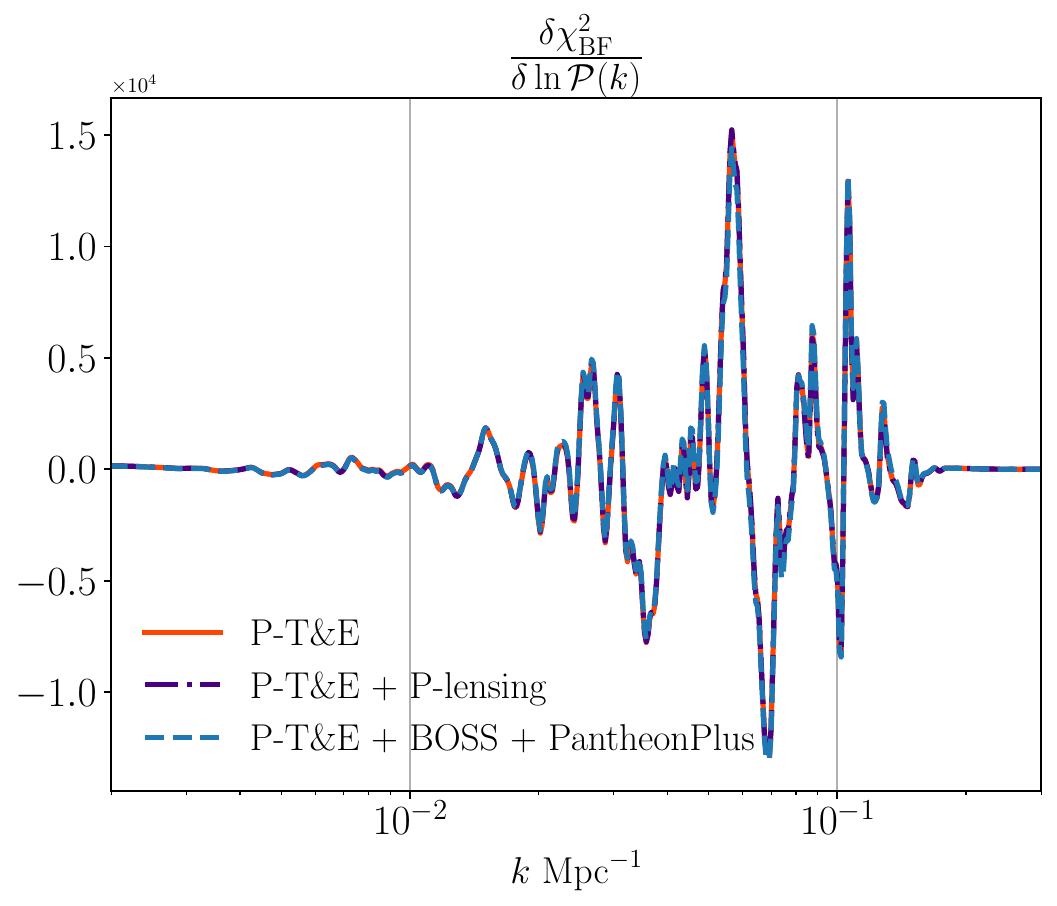}
\includegraphics[width = .33\linewidth,trim= 0 0 10 20]{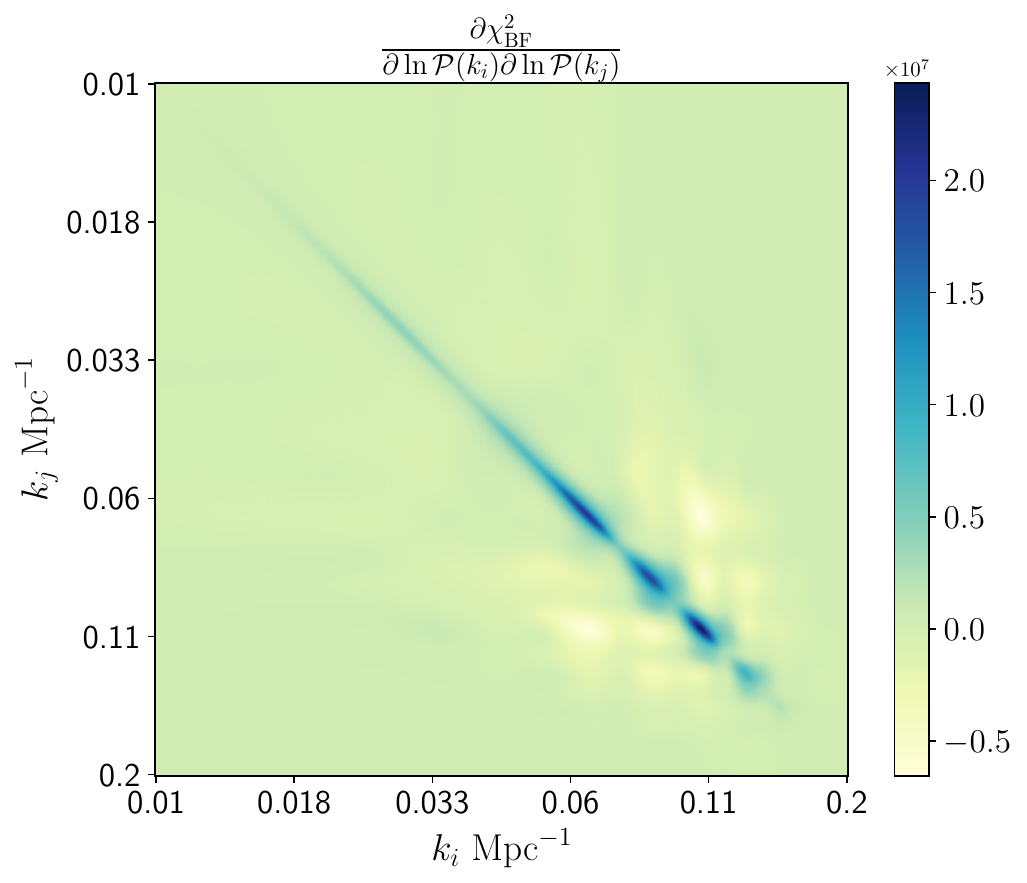}\\
\includegraphics[width = .33\linewidth,trim= 0 10 20 0]{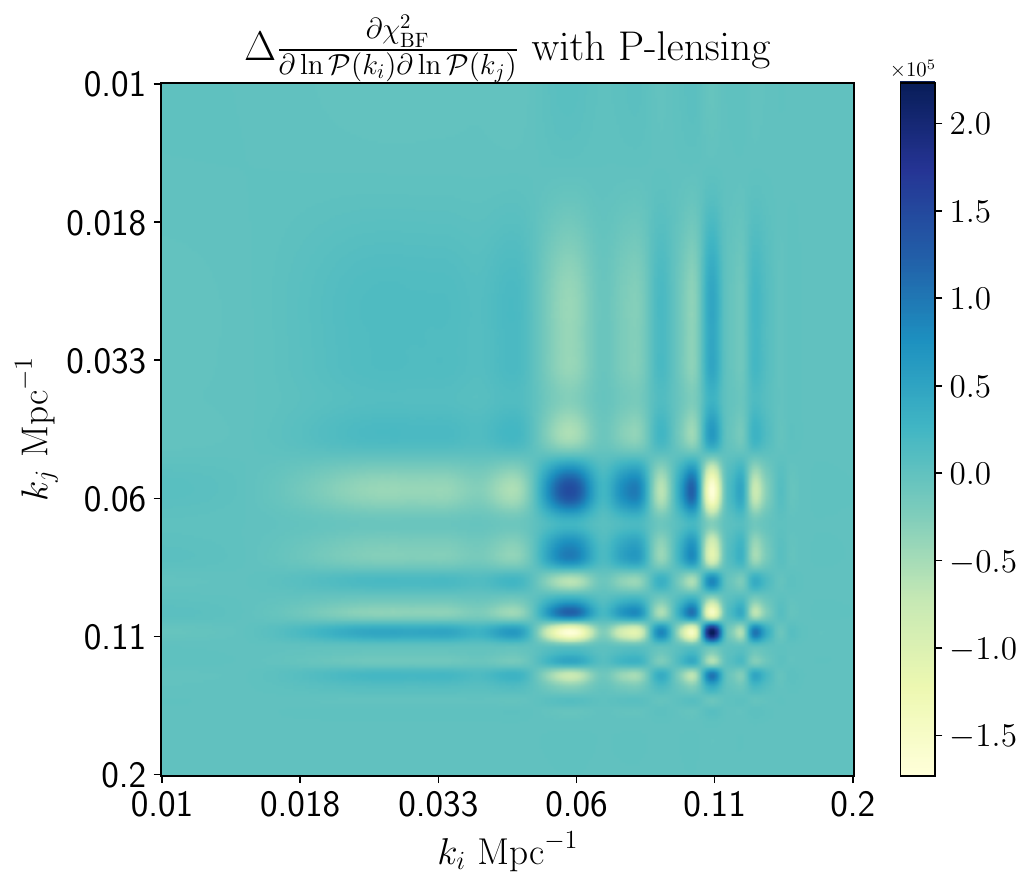}
\includegraphics[width = .32\linewidth,trim= 0 10 20 0]{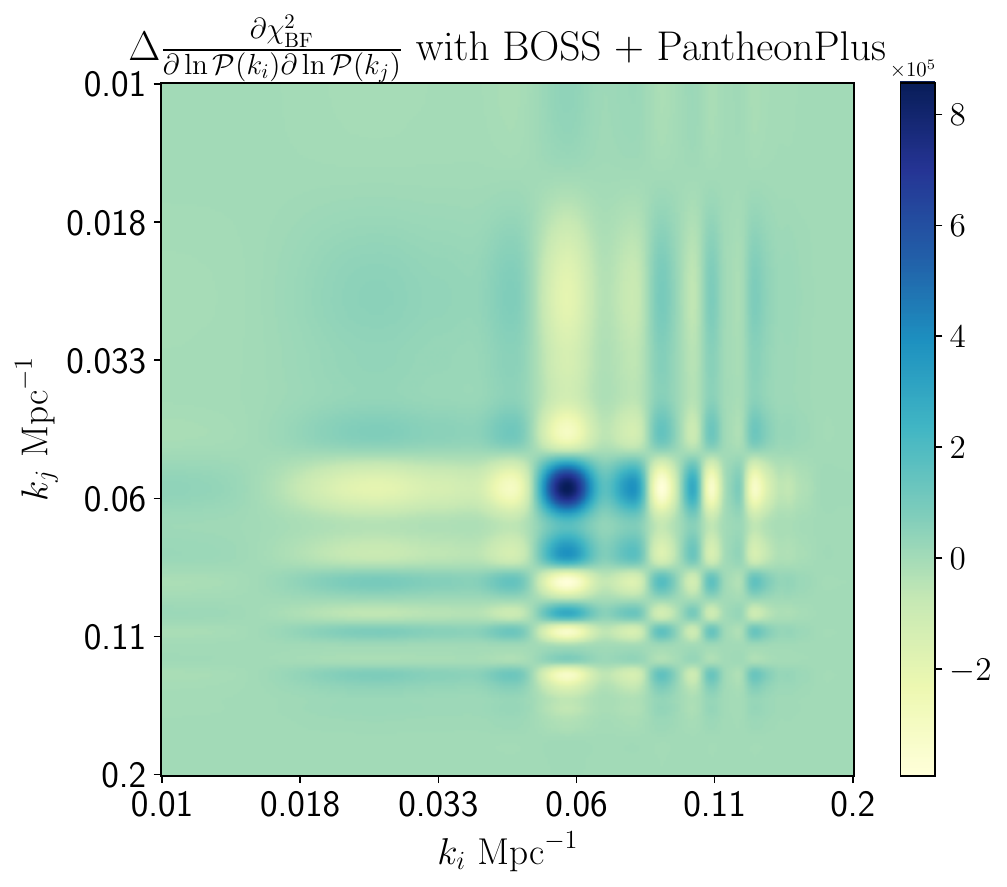}
\caption{Upper left: functional derivatives of best-fit chi-squared, Eq.~\eqref{eq:dchi2bf_lin}, given the same datasets as in Fig.~\ref{fig:deriv-Omega}. Upper right: quadratic response of a change in best-fit chi-squared, Eq.~\eqref{eq:dchi2bf_quad} with respect to logarithmic change in primordial power spectrum $\pk$ at each $k$, given Planck-T\&E data. Bottom left/right: additional contribution to the quadratic response when either Planck-lensing or BOSS/PantheonPlus data is included. Same as in Fig.~\ref{fig:deriv-Omega}, we plot with narrower range of $k$ for better presentation.}
\label{fig:deriv-chi2}
\end{figure*}

We show the functional derivatives [Eq.~\eqref{eq:dObf}--\eqref{eq:dchi2bf_quad}] in Fig.~\ref{fig:deriv-Omega} and \ref{fig:deriv-chi2}. Note that while we calculate the functional derivatives as functions of $\xi=k\eta_0$, we show them as functions of $k$ in this appendix for a better presentation.

\pagebreak

\section{Comparison with a solution for CMB lensing anomaly in Hazra et al.~2022 \cite{Hazra:2022rdl}}
\label{appendix:comparison}

\begin{figure}[h!]
\includegraphics[width = .95\columnwidth,trim= 10 20 10 0]{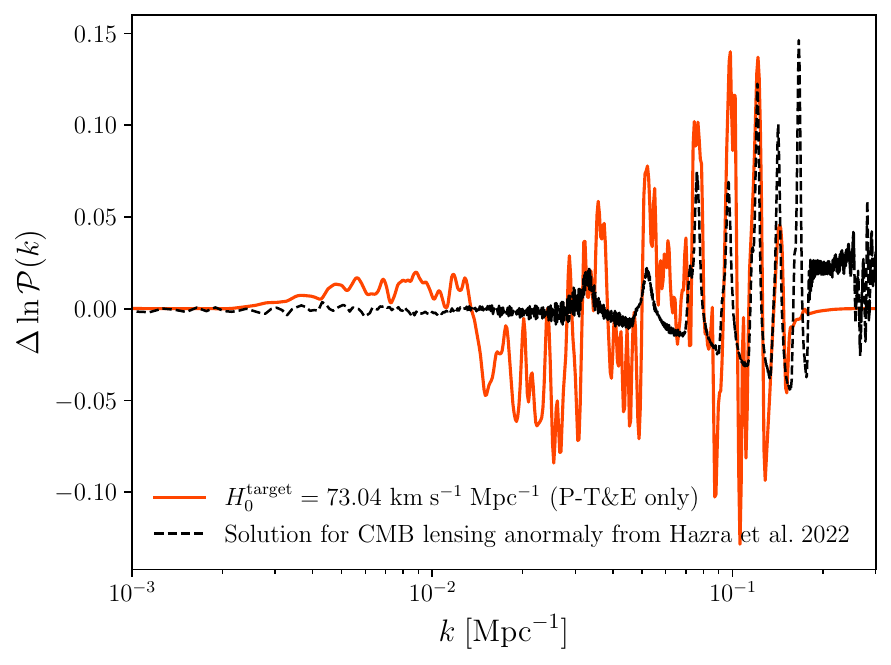}
\caption{Orange: The logarithmic modifications to the primordial power spectrum constructed to result in $H_0=73.04\;\text{km\;s}^{-1}\text{Mpc}^{-1}$ with Planck-T\&E data (same as the orange curve in the top panel of Fig.~\ref{fig:CMB-CMBBAOPantheon}). Black: A solution for Planck CMB lensing anomaly proposed in Ref.~\cite{Hazra:2022rdl}. Two modifications share similar features around $k\sim0.1/{\rm Mpc}$.}
\label{fig:sol_lensing}
\end{figure}

Interestingly, previous studies investigating modifications to the primordial power spectrum to address the CMB lensing anomaly found a modification with a similar shape on scales around $k\sim0.1/{\rm Mpc}$ to the one we identified for resolving the Hubble tension using Planck CMB spectra~\cite{Hazra:2022rdl,Antony:2022ert}. This proposed solution for CMB lensing anomaly from Ref.~\cite{Hazra:2022rdl} is over-plotted in Fig.~\ref{fig:sol_lensing} as a black curve together with our solution (orange). This could suggest that the ability to find our solution with Planck-T\&E data is at least partially influenced by the presence of the CMB lensing anomaly rather than being a fully independent resolution to the Hubble tension. Since ACT data show no evidence of excess lensing in the CMB power spectra, it would be interesting to investigate how incorporating ACT impacts our solutions, given that they seem tied to the Planck CMB lensing anomaly.

\end{appendix}

\bibliography{mybib}
\end{document}